\documentclass[prb,aps,twocolumn,showpacs,floatfix,longbibliography]{revtex4-2}
\usepackage{amsmath, amsfonts, amssymb}
\usepackage{graphicx,epsfig}
\usepackage{graphicx}
\usepackage{color}
\usepackage{siunitx}
\usepackage{hyperref}
\newcommand{\ignore}[1]{}

\begin{document}
	

\title{A low-frequency, high-amplitude, torsional oscillator \\for turbulence studies in quantum fluids}

\author{A. M.\ Gu\'enault$^{\dag}$}
 \affiliation{Department of Physics, Lancaster University, Lancaster LA1 4YB, United Kingdom}

\author{P. V. E.\ McClintock}
 \affiliation{Department of Physics, Lancaster University, Lancaster LA1 4YB, United Kingdom}

\author{M.\ Poole}
 \email{m.poole@lancaster.ac.uk}
  \affiliation{Department of Physics, Lancaster University, Lancaster LA1 4YB, United Kingdom}

\author{R.\ Schanen}
\affiliation{Department of Physics, Lancaster University, Lancaster LA1 4YB, United Kingdom}

\author{V.\ Tsepelin}
 \affiliation{Department of Physics, Lancaster University, Lancaster LA1 4YB, United Kingdom}

\author{D.\ Zmeev}
 \affiliation{Department of Physics, Lancaster University, Lancaster LA1 4YB, United Kingdom}

\author{D.\ Schmoranzer}
 \affiliation{Faculty of Mathematics and Physics, Charles University, Prague, Czech Republic}
	
\author{W. F.\ Vinen}
\affiliation{School of Physics and Astronomy,  University of Birmingham, B15 2TT , United Kingdom}

\date{\today}

\begin{abstract}
	
We describe a new type of torsional oscillator, suitable for studies of quantum fluids at frequencies of $\sim$ \SI{100}{Hz}, but capable of reaching high velocities of up to several cm\,s$^{-1}$. This requires the oscillator amplitude to exceed \SI{100}{ \mu m}, which is much too large for a conventional capacitor-driven device. We describe the new geometry for the oscillator, discuss its design, and report our initial tests of its performance.

\vspace{0.2cm}
\noindent $^{\dag}$Deceased October 30, 2019
	
\end{abstract}

\maketitle

\section{Introduction}
\label{sec:intro}

Torsional oscillators have been a valuable tool in many studies of quantum fluids and solids. They have proven to be successful tools for detecting  phase transitions in liquid helium \cite{Andronikashvil:46} and isotopic mixtures \cite{ Golov:98}, in 2D adsorbed films and in restricted geometries \cite{ Zhelev:17,Nyeki:17} and porous media \cite{ Berthold:77, Syvokon:00,Golov:04,Taniguchi:07}. The most direct experiment searching for the supersolid phase in
bulk solid $^{4}$He (performed by Bishop, Paalanen and Reppy) also used the torsional oscillator technique. The onset of superfluidity
in the helium inside the torsion oscillator decreases I the moment of inertia, and hence decreases the resonant period. Bishop et al. made measurements of solid
helium from \SI{25}{bar} to \SI{48}{bar}, and concluded that if there is a supersolid state, then either the supersolid fraction (the fraction of $^{4}$He atoms participating in superflow) is less than 5 x 10$^{-6}$ or the critical velocity is less than \SI{5}{\mu m s^{-1}} (The critical velocity is the maximum velocity of superflow without any detectable dissipation.)\cite{Bishop:81}. This same technique was also used in the detection of an anomalous response in solid helium by E. Kim and M. Chan at Pen State university \cite{Kim:04a}. Others soon followed with other designs and experiments \cite{Rittner:09,Pratt:11,Zmeev:12,Brazhnikov:12,Kim:14,Nichols:14,Fear:16}. Further studies on quantum turbulence and the study of Kelvin wave requires that we lower the frequency region of operation and increase the velocity range of the oscillator. We report here a novel design that allows us to fulfil this goal.

A common design is illustrated in Figure~1. The oscillator, shown in yellow, is composed of a cylindrical body and a support rod. The body contributes to the total moment of inertia~$I$ and the support rod provides the spring constant~$k$. To drive the oscillator and measure its motion, electrostatic activation and detection two capacitors are used. Each capacitor is formed of a moving electrode attached to the main body shown in blue, and a fixed electrode, shown in red. Each pair of electrodes is separated by a very small gap on the order of \SI{100} {\mu m}. 

\begin{figure}[t]
	\includegraphics[width=\linewidth]{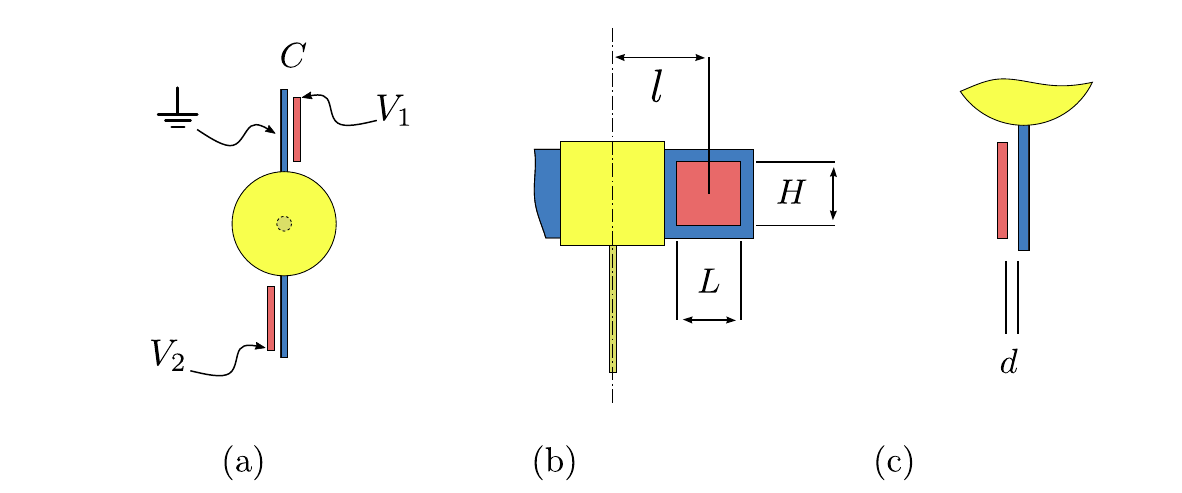}
	\caption{Common design for a torsional oscillator.
		(a) Top view showing the main body of the oscillator, the moving electrodes in blue and the fixed electrodes in red. (b) Side view of the oscillator showing the main body of the oscillator, the torsion rod, the positions~$l$ and size~$L \times H$ of each fixed plate. (c) Detail of the top view showing the gap~$d$ between the plates of one capacitor.}
	\label{fig.StandardDesign}
\end{figure}

The capacitors are both charged to a substantial DC voltage, so that at rest the plates strongly attract. An additional AC potential is then applied to the driving capacitor, which causes small oscillations of the moving plate and the oscillator body to which it is fixed. The corresponding capacitance changes in the second transducer cause a small induced AC current, measured by a phase-sensitive detector referenced to the AC drive. This system works well at small amplitudes, but is constrained by the very narrow plate separation needed in the capacitors. In recent work on superfluid $^{4}$He, a need has arisen to be able to study quantised vortex line production at high velocity, but low frequency, thus requiring amplitudes of order\SI{390}{\mu m} which corresponds to \SI{300}{mm s^{-1}} at a frequency of \SI{123.3}{Hz} needed to get to the predicted critical velocity of approximately \SI{300}{mm s^{-1}} in the superfluid filled cell. This is impossible using the conventional oscillator design because the capacitor plates would be touching and earlier conventional designs were looking at velocities of up to \SI{200}{\mu m s^{-1}} three orders of magnitude smaller than required. To solve the problem we use a quite different transducer geometry: the drive and detect capacitors have {\it partially} overlapping electrodes which are parallel to the motion of the body, not perpendicular. Forces are therefore generated in the plane of the capacitor plates; and the capacitance changes with the area of overlap, leaving the small distance between the plates unaffected. We use the same external electrical drive and detect systems as for the conventional oscillator \cite{Richardson:22a}. This newly designed oscillator can achieve velocities of up to \SI{150}{mm s^{-1}}, with stable and reproducible vacuum data and with the cell full of  $^{4}$He. In section 2 we describe the design and construction of our new oscillator, which is intended for studies of dissipation in superfluid $^{4}$He at millikelvin temperatures. In section 3 we discuss operation of the torsional oscillator. In section 4 we discuss tests of its performance in vacuum and in superfluid helium, and in section 5 we draw conclusions.

\begin{figure}[t]
	\includegraphics[width=\linewidth]{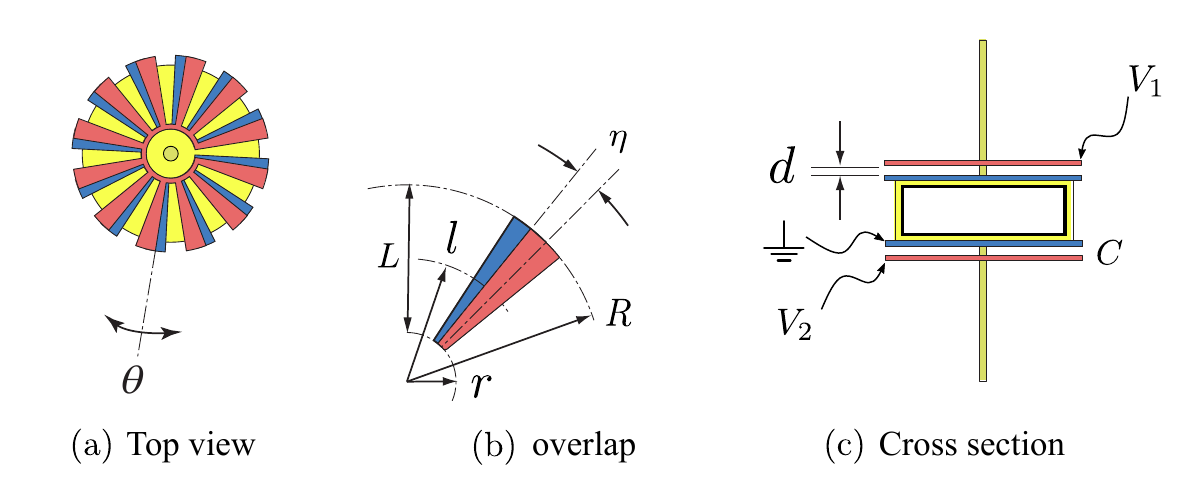}
	\caption{Capacitance geometry for a torsional oscillator with circular electrodes. (a) Top view of the oscillator. The moving electrodes are shown in blue. The fixed electrodes are shown in red. (b) Detail of the overlap geometry of electrodes. The angle~$\eta$ is the overlapping angle between a pair of facing electrodes. (c) Cross section of the oscillator showing the torsion rod and the gap between the capacitors plates.}
	\label{fig.CapacitanceDesignCircular}
\end{figure}

\section{Design and construction}
\label{sec:design}
Our need for a low-frequency \cite{Bradley:12b}, high-amplitude torsional oscillator led us to the novel geometry shown in Figure 2. Our aim is to measure critical velocities by monitoring the dissipation of the oscillator and its dependence on the amplitude of oscillation. As in a conventional torsional oscillator, the active elements are parallel plate capacitors charged by a high DC voltage, and with an AC voltage superimposed in the case of the driving electrodes. However, we organise the electrodes so that there is incomplete overlap between the plates of each capacitor. When the DC voltage is applied, there is a force tending to increase the overlap and to bring the plates towards alignment. To make full use of this effect, and to spread the force evenly around the torsional oscillator, we developed the design illustrated in Figure 2(a). Each electrode has 12 evenly-spaced sectors, electrically connected together, manufactured using conventional lithographic techniques on a thin circular circuit board of radius 2 cm. The sample of liquid $^4$He for investigation is placed in a pillbox-shaped epoxy cell fixed to the centre of a rigid copper framework by a BeCu tube, which forms the torsion rod. This tube is attached rigidly to the top and bottom of the framework, and it has a small hole in the side at the centre so that it also acts as the fill line for the cell. This highly symmetrical design helps to ensure that the oscillator does not respond to any unwanted nuisance modes at nearby frequencies. Two finned electrodes of the type described above are permanently fixed to the top and bottom of the cell. Very close to them, two adjustable electrodes are positioned above and below, with the deliberate radial misalignment illustrated in Figure 2(a). These are attached to the frame and carefully adjusted to create a narrow gap $d$ for each capacitor, one of which is used to drive the oscillator and the other to detect its motion. The full assembly is shown in Figure 3.

\section{Circular electrodes and operating Torsional oscillator}
\label{Circular electrodes and operating Torsional oscillatorr}

 Each electrode consists of a set of smaller radial segments distributed in a star like fashion and connected by a small ring. The capacitance in each case is formed by using two such plates, closely spaced, and shifted in angle by one half-segment-width, as shown in Figure 2(b). Its capacitance is:

\begin{equation}
	C = \epsilon_0\frac{S}{d} =
	\epsilon_0\frac{\pi(R^2 - r^2)}{d}\times\frac{\eta}{2\pi}=
	\epsilon_0\frac{R^2 - r^2}{2d}\eta =
	\epsilon_0\frac{L l}{d}\eta
\end{equation}
where~$L=R-r$ is the length of one electrode, $l=(R+r)/2$ the position of its centre from the axis of rotation and~$\eta$ the angle of overlap between the two electrodes. For~$N$ electrodes in parallel, we find the total capacitance is:
\begin{equation}
	\label{eq.circularCapacitance}
	C = N \epsilon_0\frac{L l}{d}\eta \,.
\end{equation}

In contrast to the conventional geometry, the narrow gap between the electrodes remains fixed, and the change of capacitance arises on account of variations of the overlap angle during oscillation. With 12 electrodes, $R$ =20\,mm, $r$ = 7.5\,mm giving $L$ = 12.5\,mm and $l$ = 13.75\,mm with a gap of \SI{100}{\mu m} and an overlap of $6 \deg$, we expect a capacitance of \SI{19.12}{\pF}.

\begin{figure}[t]
	\includegraphics[width=\linewidth]{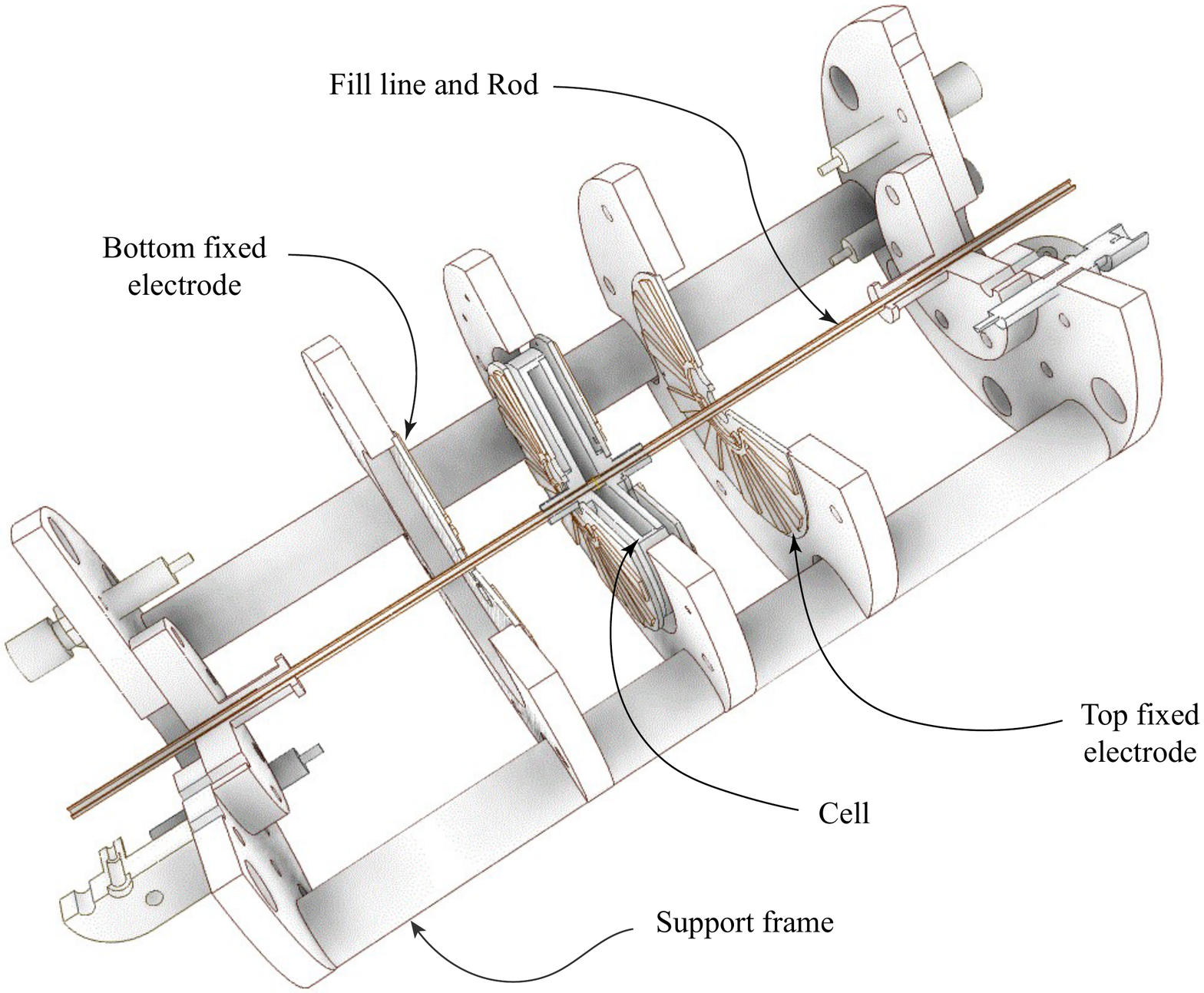}
	\caption{Drawing of the assembly, showing the torsional oscillator with circular electrodes and its surrounding frame.The pill-box-shaped helium cell is in the centre, and the fixed electrodes have been pulled away from it for clarity though, in reality, they are positioned very close to it as shown in Fig. 2(c).}
	\label{fig.Full Assembly of new design}
\end{figure}

\subsection{Energy and torque}

For a potential difference~$V$, the electric energy stored in the capacitor is
\begin{equation}
	U = \frac{1}{2} C V^2 = \frac{1}{2} \epsilon_0 \frac{N L l}{d} \eta V^2 \,.
\end{equation}
which is almost identical to that for the conventional torsional oscillator which is given by
\begin{equation}
	U = \frac{1}{2} C V^2 = \frac{1}{2} \epsilon_0 \frac{ L H}{d}  V^2 \,.
\end{equation}
where $L H$  is the area of the electrode:

The torque being applied to the oscillator is
\begin{equation}
	\label{Eq.OscDrive2}
	\tau_1 = \frac{1}{2}\epsilon_0 \frac{N L l}{d} V^2 \,.
\end{equation}
The torque always tends to increase the overlap between the capacitor plates. The torque does not depend on the sign of the potential $V$.

\subsection{Static displacement}

In reaction to an angular displacement~$\theta$ at a radius of \SI{0.8}{mm} of the torsion rod, the rod exerts a torque~$\tau_2 = -\kappa\theta$ leading to the equilibrium position

\begin{equation}
	\theta = \frac{1}{2}\epsilon_0 \frac{N L l}{\kappa d} V^2
\end{equation}
 Here kappa is the spring constant. For a static potential of $200$ V the displacement of the inner radius $r$ of the electrodes is $x = \theta l = 16.7$\,nm

\subsection{Harmonic drive}
 The force is parabolic in~$V$. If we linearise the force by adding a fixed voltage~$V_S$ much larger than~$V$ we can increase the driving force and increase the sensitivity of our oscillator in the linear regime. We find the amplitude of the harmonic part of the driving torque

\begin{equation}
	\label{Eq.drivetorque}
	\tau = \epsilon_0\frac{NLl}{d}V_S V
\end{equation}
which is linear in~$V$. We then get the zero frequency response with the AC drive

\begin{equation}
	\theta_0 = \epsilon_0 \frac{N L l}{\kappa d} V_S V \,.
\end{equation}
Using \SI{5}{V} for the drive we expect the displacement amplitude of the electrode to be $x_0 = \theta_0 l = 0.84$\.nm. This is about 8~times less than for the conventional oscillator design.

\subsection{Harmonic displacement and velocity}
From the oscillator amplitude at resonance we get

\begin{equation}
	\theta = Q \times \theta_0 = Q \times \frac{\epsilon_0 N L l}{\kappa d} V_S V
\end{equation}
and hence the oscillator velocity

\begin{equation}
	\label{Eq.circularvelocity}
	\dot\theta
	= \omega_0 \times \theta
	= \omega_0 \times Q \frac{\epsilon_0 N L l}{\kappa d} V_S V \,.
\end{equation}
This gives an amplitude of~$x = l \theta$ =2.09 $\mu$m and a velocity ${\dot x} = l {\dot \theta}$ = 1.97\.mm s$^{-1}$.

\subsection{Detection}
From equation(\ref{eq.circularCapacitance}), we get the detection current

\begin{equation}
	i = \dot C V_D = \frac{\epsilon_0 N L l}{d} V_D \dot\eta \,.
\end{equation}
Because ${\dot \theta} = {\dot \eta}$, we can express the oscillator angular velocity $\dot \theta$ in terms of the voltage output~$u$:

\begin{equation}
	\dot \theta = \frac{d}{\epsilon_0 N L l V_D} \frac{u}{G} \,.
	\label{eq.ThetaDotFromSignal}
\end{equation}

\subsection{Measured velocity (from signal)}
In practice, however, we do not need to measure the gap in order to determine the velocity. If we replace $d / \epsilon_0 N L l$ by $\eta / C$ we see that the velocity of the oscillator at radius~$R$ is

\begin{equation}
	v  = R \dot{\theta} = R\frac{\eta}{C V_D} \frac{u}{G}
\end{equation}
which is independent of $d$. All quantities are known or can be measured accurately, except for $\eta$. Therefore we can expect to calculate the velocity from the measured signal to within 12\% as discussed in Section IV.B.

\subsection{Expected velocity (from drive)}
Knowing the driving force, we can calculate the velocity to be expected from (\ref{Eq.circularvelocity}). Replacing $\epsilon_0 N L l / d$ by~$C /\eta$, we get

\begin{equation}
	v = R \left( \frac{\omega_0 Q}{\kappa} \right) \frac{C}{\eta} V_S V
\end{equation}
where the value of $\kappa$ is not known accurately. However, the moment of inertia can be calculated with some certainty. From $\omega^2_0 = k/I$ where $I$ is the moment of inertia of the oscillator, we can rewrite as

\begin{equation}
	v = R \left( \frac{Q}{\omega_0 I} \right) \frac{C}{\eta} V_S V \,.
\end{equation}

\subsection{Oscillator constant HW/D}
The oscillator constant, the so called ``height-times-width over drive" (HW/D), can be calculated theoretically where the height is the signal output $u$ in $V$, the width is the resonance linewidth $\Delta f = \Delta\omega/2\pi$ in [Hz] and the drive is $V$. This easily measured constant should be characteristic of the oscillator geometry and independent of the damping factors. The ``height-times-width over drive" does not change with drive or frequency, whether at room temperature or base temperature; or with the cell empty or full of helium. This constant should be the same value and is a useful check for consistancy under all experimental conditions. From (\ref{Eq.circularvelocity}) and (\ref{eq.ThetaDotFromSignal}) we have respectively 

\begin{equation}
	u = G\frac{C_1}{\eta}V_D\dot{\theta}
\end{equation}
and
\begin{equation}
	\dot{\theta} = G\frac{Q}{\omega_0 I}\frac{C_2}{\eta}V_S V \,.
\end{equation}
where~$C_1$ and $C_2$ are respectively the detection and drive capacitances. Replacing~$\theta$ and using~$Q=\omega_0/\Delta\omega$ gives
\begin{equation}
	u\frac{\Delta f}{V} =
	\frac{1}{2\pi I} \times \frac{C_1 C_2}{\eta^2} \times G V_D V_S
\end{equation}
with all quantities on the right hand side either known or independently measurable.

\section{Experimental Tests}
 During the development of the oscillator, we made several iterations and re-designs to eliminate problems that arose. In each case, the oscillator was first tested at room temperature. In doing so, we discovered that the circuit board forming the substrate of the electrodes is not electrically inert at room temperature. Over a few hours, a high voltage gradually polarises the material, and so reduces the signal height at resonance.  In our case, humidity (water content in the plastic) could be responsible for the residual conduction; but at nitrogen temperature the problem does not arise or happens so very slowly that it appears to be undetectable over the life time of the experimental run. For all subsequent experiments, therefore, we ensured that we held $V$ = $0$ for a day before cooling. With the development of the torsional oscillator we also used different diameter torsion rods made from BeCu and sterling silver. In this paper we present test data from 2 cells with BeCu torsion rods. The data at \SI{79.1}{\Hz} is a 1.2 mm od and 0.9 mm id BeCu torsion rod while data at \SI{123.3}{\Hz} is a 1.6 mm od and 1.24 mm id BeCu torsion rod.

\subsection{Low Temperature testing and measurements }

\subsubsection{Frequency sweeps}
The oscillator was installed in a dilution refrigerator, with the fill line and the copper framework (Figure 3) thermally anchored to the mixing chamber. We have completed some preliminary tests of the oscillator in vacuum and with liquid helium inside the experimental cell. The capacitor plates were set as close as possible while avoiding touches. The capacitances of the upper and lower electrode pairs were measured as $11.1$\,pF and $11.3$\,pF respectively. Initial indications showed that the device was working well with the resonance frequency $79.106$\,Hz, quality factor $Q$ $\sim 37,800$ at base temperature of dilution refridgerator.

With the cell filled with liquid helium at zero pressure and a temperature of about 20 mK, we made a sequence of frequency sweeps with an AC driving amplitude of $2.0$ $V$\.rms. The DC supplied to both drive and detect capacitors was $200$\,$V$ as shown in Figure 4. These measurements were made over 14 hours with each sweep taking 10,000 seconds, demonstrating that the stability of the oscillator is very good at low temperatures.

\begin{figure}[h]
	\includegraphics[width=\linewidth]{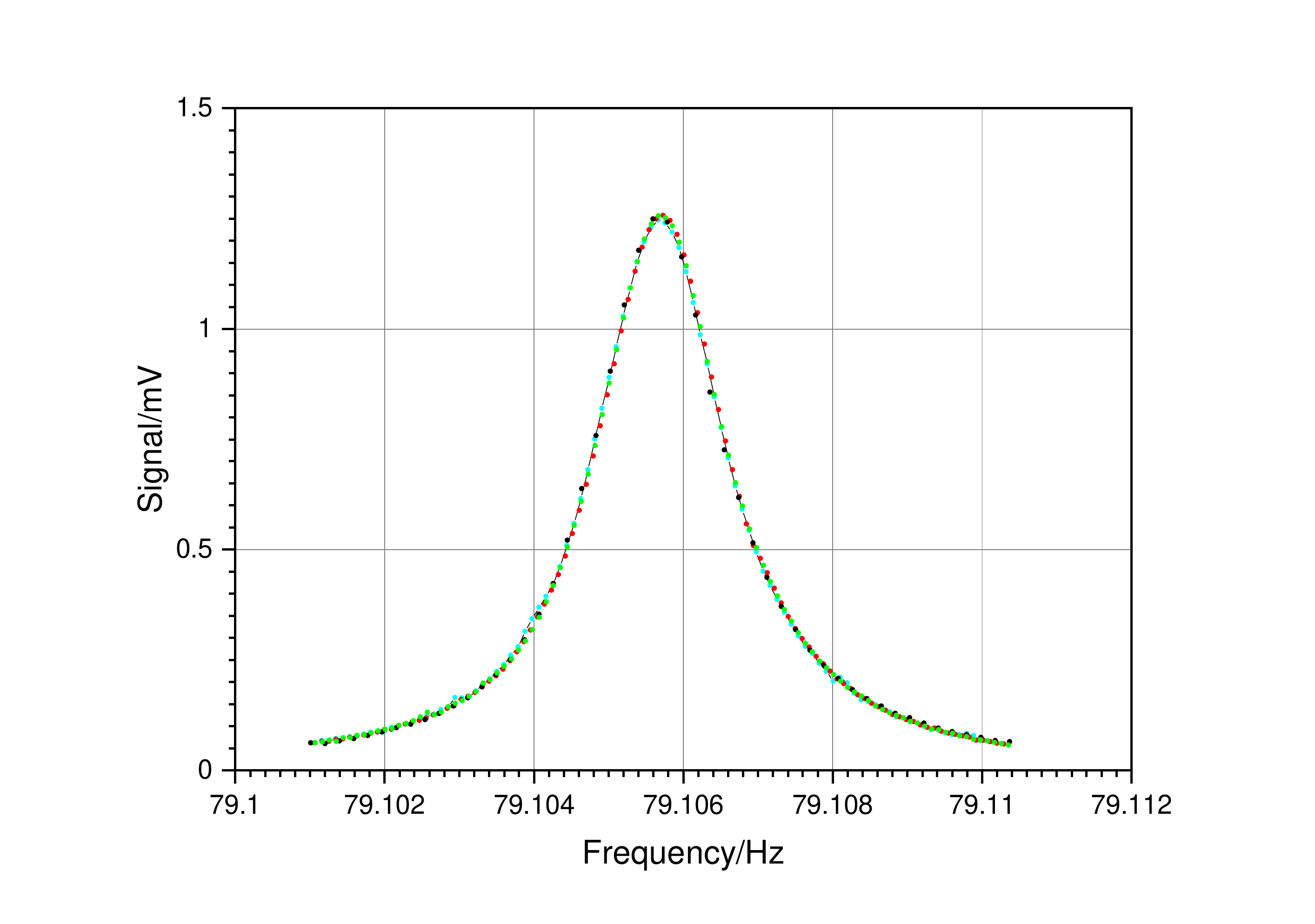}
	\caption{Frequency sweeps at base temperature with superfluid  $^{4}$He in the cell. There are 5 sweeps taken over 14 hours with each sweep taking 10,000 seconds. These measurements were taken at a pressure of zero bar and a temperature of $\sim 20$\,mK}
	\label{fig.Stable frequency sweeps}
\end{figure}
The pill-box geometry ensures that there is no flow over a convex surface, thus eliminating a common form of
classical instability.
We took a series of measurements at different DC and AC drive voltages both at vacuum  and with the cell full of $^{4}$He. One set of data taken with various AC voltages both in vacuum and with helium in the cell are a set of frequency sweeps, (again taken over several hours at base temperature) and these are shown in Figure 5. We have converted these to show the  peak velocity in mm\,s$^{-1}$ at the outer rim of the cell. It is evident that the high amplitude sweeps are characterised by small but significant non-linear damping effects which are due to non-linear restoring forces of the torsion rod, leading to slightly asymmetric line-shapes.
Repeating the same tests with the cell filled with superfluid $^{4}$He at base temperature revealed similar slightly asymmetric but reproducible line-shapes at the higher drives.

\begin{figure}[b]
	\includegraphics[width=\linewidth]{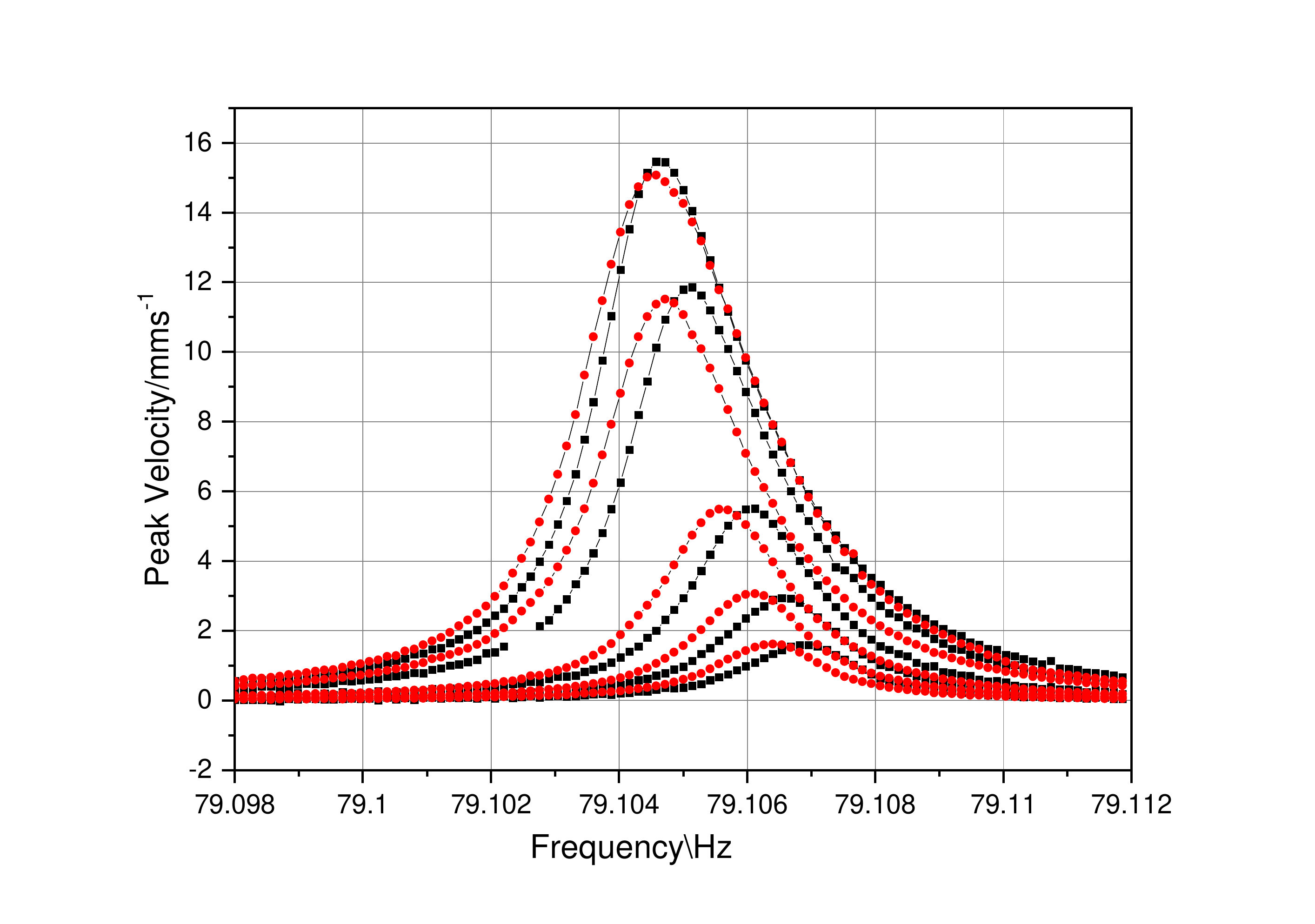}
	\caption{Frequency sweeps at several AC drives with vacuum in cell (black squares) and cell full (red circles) of superfluid $^{4}$He at base temperature.}
	\label{fig. frequency sweeps in vacuum and cell full.}
\end{figure}

We also investigated an oscillator with a larger diameter BeCu torsion rod, which we hoped would be less non-linear and able to reach higher velocities. Again, we took several different measurements with the cell at vacuum and cell full of both superfluid $^{4}$He and normal fluid $^{4}$He; again not all data is presented in this paper. At base temperature the frequency in vacuum was \SI{123.319}{\Hz} with a width of \SI{3.32}{\mHz} and a $Q$ of over 37,000. A typical resonance is shown in Figure 6. With the stiffer torsion rod there was still some non-linearity but the oscillator could reach much higher velocities and it is envisaged that with some improvements this oscillator could reach velocities of over $300$ mm\,s$^{-1}$ which are needed for the planned experiments.

\begin{figure}[t]
	\includegraphics[width=\linewidth]{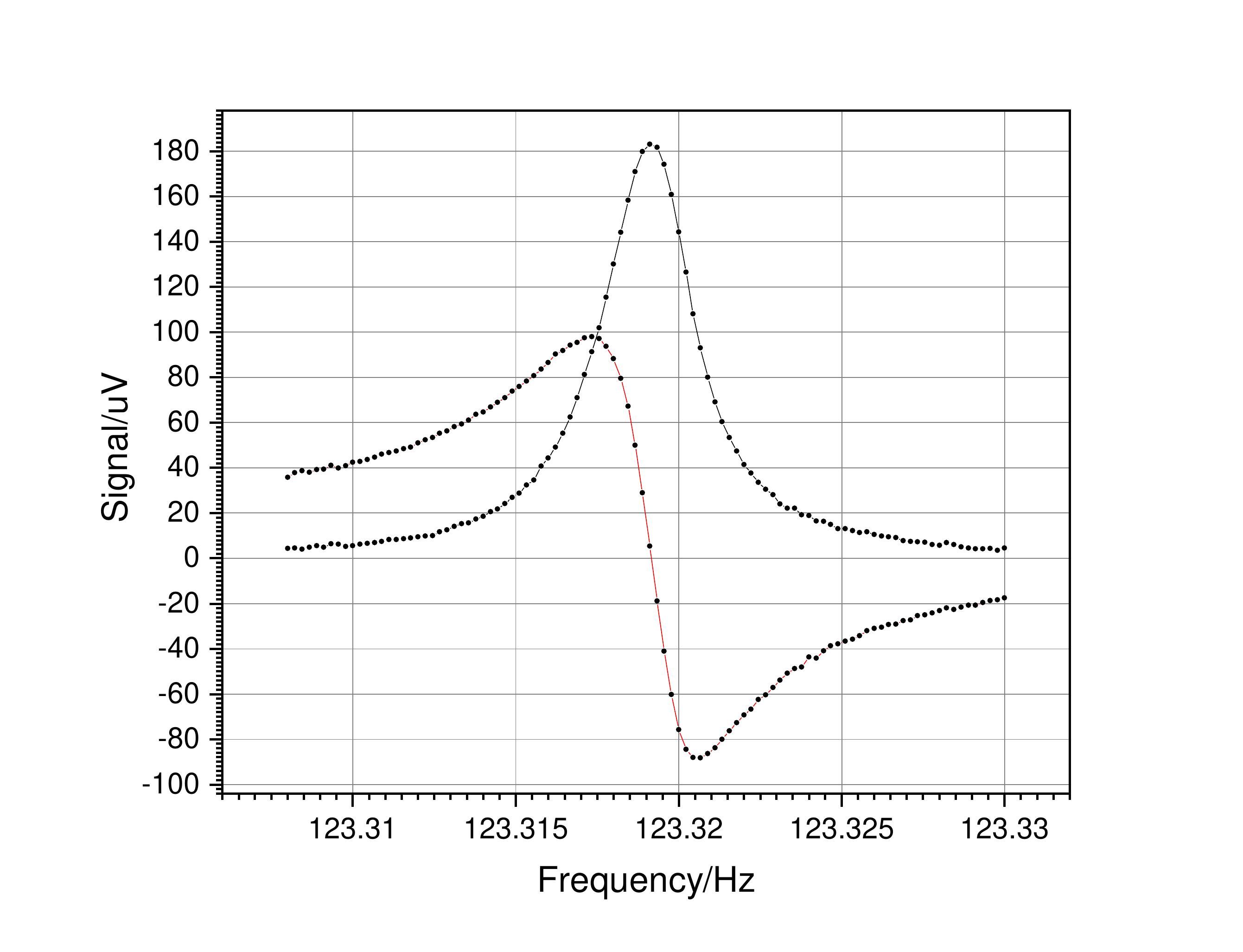}
	\caption{Frequency sweep of the oscillator with a larger diameter (stiffer) BeCu torsion rod  at base temperature in vacuum with a DC drive of 200 volts and AC drive of 125 mV.}
	\label{fig. frequency sweep.}
\end{figure}

At the higher velocities the torsional oscillator becomes non-linear and the width increases. There is a noticeable warming of the cell at these higher drives. 

\subsection{Capacitance and gap}
The only quantity that is not directly measurable is the gap~$d$ between the electrodes, but it can readily be deduced from the measured capacitance provided that we know $\eta$. The error in the overlap can be estimated. We are using centring pins of diameter \SI{0.5}{mm} at a radius of \SI{20}{mm} at the moment of fixing the electrodes. This is equivalent to $1.43\deg$. An overestimation of half a pin diameter gives an error of 12\% in the determination of the value of $\eta$. The error in reading the capacitance with the capacitance bridge is negligible.

\subsection{Amplitude Sweeps}

At base temperature we carried out amplitude sweeps both in vacuum and with the cell full of superfluid $^{4}$He. Some of the data are shown in Figure 7(a) on log-log scales. At the lower drives and velocities the data are clearly linear but, at higher velocities, all the data exhibit distinct curvature due patially to the non-linear restoring forces of the torsion rod, but there appears to be some extra damping due to the helium. Also shown Figure 7(b) is the effective damping at higher velocities. 
\begin{figure}[hb]
	\includegraphics[width=10cm]{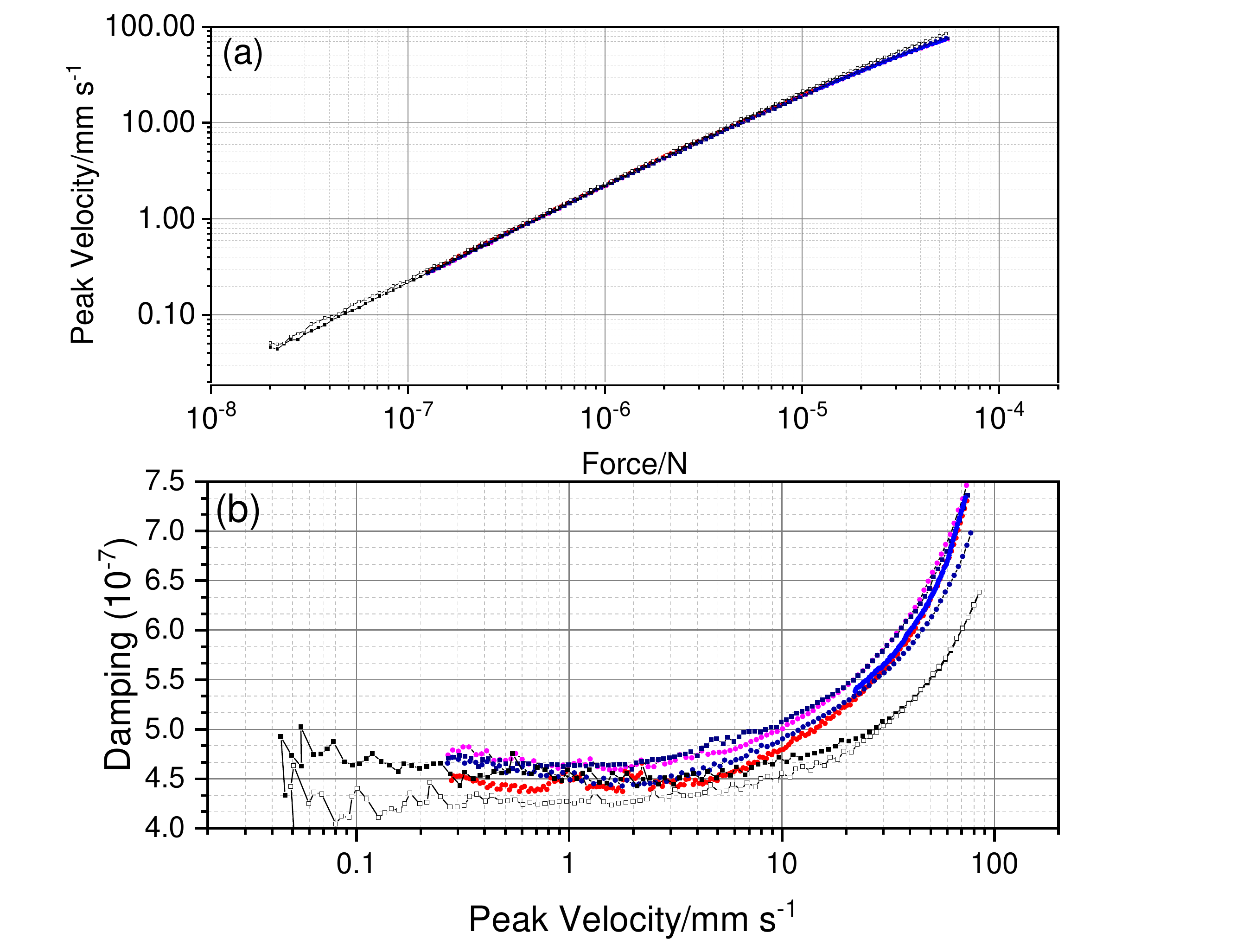}
	\caption{ Top. The peak velocity as a function of force, plotted on log-log scales. Black squares represent vacuum data. Red and blue circles are with cell full. Bottom. Shows the effective damping versus velocity. }
	\label{fig. ForcevsVelocity1.}
\end{figure}

In this version of the paper we report the values of velocity inferred from (13). However, direct optical measurements of the displacement of the oscillator at room temperature 
suggest that Eq. (13) significantly underestimates the velocity. We are continuing to search for the source of this discrepancy.

\section{Conclusion}
The ability of our oscillator to attain velocities of several cm s$^{-1}$ at such low frequencies compared to the several tens of $\mu$m s$^{-1}$ in conventional designs means that it is well suited to the studies of the nucleation of quantised vortices in superfluid $^{4}$He where critical velocities of this order are to be anticipated. Also, the decrease in sensitivity expected from the change in geometry is easily compensated by the much larger increase in the velocity that is achieved.

\section*{Acknowledgements}

We gratefully acknowledge help, advice, and valuable discussions with Oleg Kolosov in relation to the optical measurements that gave us an independent method of measuring the velocity at the rim of our cell. This work was supported by the Engineering and Physical Sciences Research Council [grant number EP/P022197/1].

\bibliography{lowtemp} 
\end{document}